# Plasmonic materials for energy: from physics to applications

Svetlana V. Boriskina, Hadi Ghasemi and Gang Chen
*Mechanical Engineering Department, Massachusetts Institute of Technology Cambridge, Massachusetts, 02139, USA*

**Abstract**

Physical mechanisms unique to plasmonic materials, which can be exploited for the existing and emerging applications of plasmonics for renewable energy technologies, are reviewed. The hybrid nature of surface plasmon (SP) modes – propagating surface plasmon polaritons (SPPs) and localized surface plasmons (LSPs) – as collective photon-electron oscillations makes them attractive candidates for energy applications. High density of optical states in the vicinity of plasmonic structures enhances light absorption and emission, enables localized heating, and drives near-field heat exchange between hot and cold surfaces. SP modes channel the energy of absorbed photons directly to the free electrons, and the generated hot electrons can be utilized in thermoelectric, photovoltaic and photo-catalytic platforms. Advantages and disadvantages of using plasmonics over conventional technologies for solar energy and waste heat harvesting are discussed, and areas where plasmonics is expected to lead to performance improvements not achievable by other methods are identified.

## 1. Introduction

Ever-growing demand for cheap, abundant and renewable energy sources fuels high interest to harvesting solar energy or waste heat from both industrial complexes and consumer appliances. All forms of light-matter interaction, which are an integral part of radiation-induced energy conversion processes, are heavily influenced by the optical density of states (DOS) in various materials. The DOS defines the number of 'channels' for storing and/or routing the electromagnetic energy in a given medium. Nanostructured materials with DOS values higher than that of the ordinary isotropic media can be engineered, enabling enhanced light trapping and tailored emission spectra, which can be utilized in solid-state lighting, solar-thermal technologies, photovoltaic (PV), and thermophotovoltaic (TPV), and photocatalytic energy conversion schemes.

Plasmonic materials have attractive characteristics for many such applications due to their unprecedented sub-wavelength light focusing capability and DOS modification. The unique properties of plasmonic materials stem from the resonant collective oscillations of charge carriers – so-called plasmons – which can be induced by external light sources, embedded emitters or thermal fluctuations [1-3]. The collective dynamics of volume plasmon oscillations is driven by the long-range Coulomb forces, and can be controlled by tailoring the spatial region filled by electron plasma. In particular, plasmons confined to metal-dielectric interfaces or thin metal films couple with photons resulting in the formation of surface plasmon-polariton (SPP) waves [4, 5]. SPPs feature flat dispersion characteristics with wavelength much shorter than the free-space propagating waves, which has been explored for information processing [1, 6, 7]. On the other hand, the flat dispersion also results in large DOS at select frequencies corresponding to the excitation of SPP modes. In turn, wavelength- and sub-wavelength-scale particles support quantized localized surface plasmon (LSP) modes [8, 9] with a large DOS.

The ability of plasmonic materials and particles to generate large a DOS in their vicinity has been successfully used for bio(chemical) sensing, fluorescence and Raman spectroscopy [10-16], and is particularly interesting for energy applications. The energy stored in the high-DOS SP oscillations can be coupled to propagating waves, to other SP modes, to electrons or converted into heat. These high DOS characteristics pave the way to a variety of potential applications of SPs, including light trapping in solar PV and TPV platforms [17-22], light extraction in solid-state lighting [23, 24], undercooled boiling [25-27], enhancement of frequency up-conversion processes [28], thermoelectric energy conversion [29], nanoscale heat management [30, 31], hot-electron PV cells and photo-detectors [32-35], near-field thermophotovoltaics [36], water splitting to produce hydrogen through photocatalytic reactions [37-40], 'smart' energy-saving window coatings [41, 42], and optical data storage [43]. In this article, we will discuss how the large DOS, heat, and hot electrons generated inside and in the vicinity of plasmonic nanostructures can be harnessed for energy applications.

## 2. Surface-plasmon-mediated density of states modifications

The DOS is a very important characteristic of the material as it defines the number of channels in energy and momentum space that are available for storing electromagnetic energy and interacting with external fields in the absorption and emission processes. The electromagnetic energy density per unit volume per unit energy in a material is defined via the well-known Planck formula as: $U = \hbar\omega \cdot f(\omega) \cdot D(\omega) d\omega$, where $\hbar\omega$ is the photon energy, $f(\omega) = \left(\exp\left((\hbar\omega - \mu)/k_B T\right) - 1\right)^{-1}$ is the Bose-Einstein statistics ($k_B$ is the Boltzmann constant, $T$ is the temperature, and $\mu$ is the chemical potential, $\mu = 0$ for thermal radiation), and $D(\omega)$ is the DOS [44]. A specific form of DOS as a function of photon energy can be derived from the general expression for DOS in the momentum space $D(k) = d\Omega \cdot k^2/(2\pi)^3$ if the dispersion relation of the material is known ($\mathbf{p} = \hbar\mathbf{k}$ is the photon momentum, $\mathbf{k}$ is the wavevector, $k = |\mathbf{k}|$, and $d\Omega$ is the unit angular range to which the wavevectors are confined).

The dispersion relation for an isotropic bulk dielectric or semiconductor with refractive index $n$ (Fig. 1(a1)) has a well-known form $k_x^2 + k_y^2 + k_z^2 = n^2\omega^2/c^2$, which is shown in Fig. 1(a2). In this case, the wavevectors of the optical states available at each frequency are confined to the surface of the sphere (Fig 1(a3)). Accordingly, the DOS of isotropic bulk dielectric for both light polarizations is $D(\omega) = \omega^2 n^3/\pi^2 c^3$. As schematically illustrated in Fig. 1(a4), this relatively low DOS is a limiting factor in manipulating and enhancing optical absorption and emission processes in bulk dielectrics or semiconductors [45-47].

Plasmonic materials can provide large DOS at select frequencies, which results from the availability of high-momentum states not accessible in ordinary materials. The dielectric permittivities of plasmonic materials are defined by a Drude model $\varepsilon_m(\omega) = \varepsilon_\infty - \omega_p^2/(\omega^2 + i\gamma\omega)$, where $\varepsilon_\infty$ is the high-frequency permittivity limit, $\omega_p^2$ is the plasma frequency and $\gamma$ is the electron collision frequency. What sets plasmonic materials aside from dielectrics is the fact that the real part of the Drude permittivity becomes *negative* below a certain frequency. We will show in the following that the negativity of dielectric constant has far-reaching implications for the spatial and spectral control of DOS in plasmonic structures and materials.

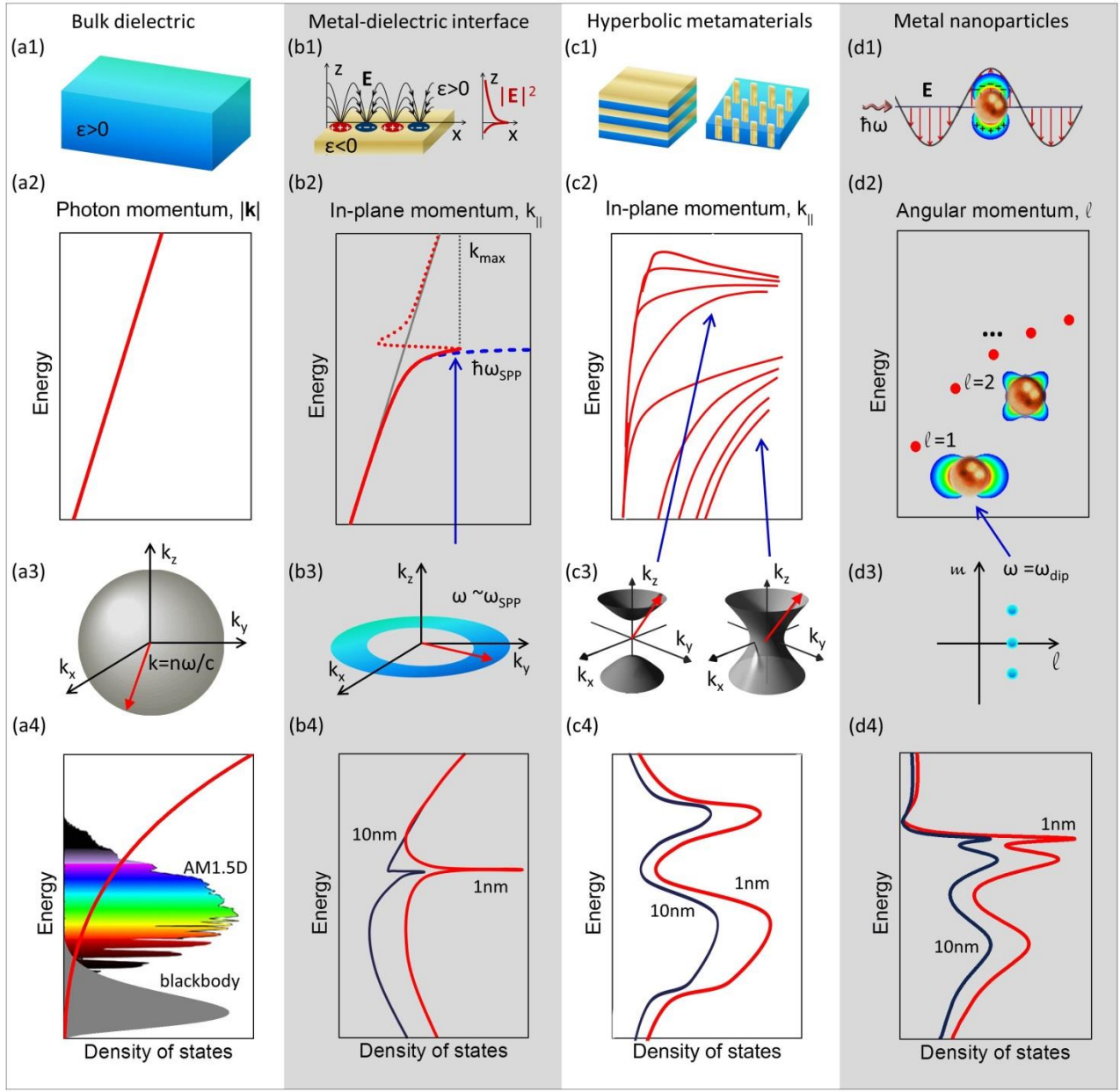

**Fig. 1. Evolution of the dispersion characteristics and DOS in nanostructured plasmonic materials**. (a1-d1) Schematics of materials and nanostructures with various degrees of light and charge carriers confinement, including (a1) bulk isotropic dielectric, (b1) a metal-dielectric interface, (c1) thin-film-based and wire-based hyperbolic metamaterials (metal inclusions are golden-colored), and (d1) a spherical metal nanoparticle. (a2-d2) The dispersion relations for the structures shown in (a1-d1). In (b2), the red solid line corresponds to the bound SPP mode and the red dotted line to quasi-bound and propagating modes. The dispersion relations for a lossless metal-dielectric interface (dash blue) and bulk dielectric (solid grey) are also shown. The insets to (d2) show spatial intensity distributions of dipole and quadrupole LSP modes with the angular momentum quantum numbers $l=1$ and $l=2$, respectively. (a3-d3) The momentum-space representations of the dispersion relations (iso-frequency surfaces), with allowed wavevectors shown as red arrows. In (d3), the three allowed values of the azimuthal momentum quantum number ($m=-1,0,1$) are shown for the dipole LSP mode ($l=1$). (a4) DOS in the isotropic bulk dielectric with index $n$ as a function of photon energy (red line). The energy spectra of the terrestrial solar radiation and the blackbody thermal emission are shown in the background. (b4-d4) LDOS in the vicinity of the structures in (b1-d1) as a function of energy at varying distance from the metal interface.

The simplest example of the structure that gives access to high-momentum optical states is the interface between metal and dielectric with permittivity $\varepsilon_d$ shown in Fig. 1(b1). This interface can support a surface mode – termed the surface plasmon polariton – which is a solution to the dispersion equation $k_x^{(SPP)}(\omega) = \omega/c \cdot (\varepsilon_m \varepsilon_d/(\varepsilon_m + \varepsilon_d))^{1/2}$ that only exists for $\varepsilon_m < -\varepsilon_d$ [5]. The SPP dispersion relation is plotted in Fig. 1(d) for the case of lossless (blue dash line) and lossy (red lines) metals. While for the lossless case the SPP branch extends to infinite wavevectors, the dissipative losses limit the maximum accessible value of the k-vector to $k_{max}$. Around the SPP frequency, $\omega_{SPP} = \omega_p/\sqrt{\varepsilon_\infty + \varepsilon_d}$, the available k-vectors form a ring in the momentum space (Fig. 1(b3)) with $\text{Re}(k_z^{(SPP)}(\omega)) = 0$.

The flat dispersion of the SPP mode seen in Fig. 1(b2) translates into the large DOS values at $\omega_{SPP}$, since DOS is inversely proportional to the mode group velocity $v_g^{(SPP)} = (dk^{(SPP)}/d\omega)^{-1}$. SPP being the surface wave, its intensity and DOS decay exponentially away from the interface, as shown schematically in Fig. 1(b1)). For surface waves, the local density of states (LDOS) is a more meaningful characteristic, which is plotted in Fig. 1(b4) at two points above the interface [30]. The excitation of the SPP is manifested by a sharp peak at $\omega_{SPP}$ in Fig. 1(b4). The narrow width of this peak, which depends on SPP damping, can be very useful for sensing, but for energy applications, there is a trade-off between the width and the amplitude of the peak as the total energy across the bandwidth is often desirable. Furthermore, unlike the bulk DOS, which has similar form for both light polarizations, the LDOS at the metal-dielectric interface only features the SPP-induced peak for the *p* polarization.

Significant improvements in LDOS arising from SPP confinement to the surface can be further built upon by confining plasmons to multiple thin metal films or wires (Fig. 1(c1)). Near-field optical coupling between SPP waves formed on each surface or wire results in formation of multiple SPP branches in the dispersion characteristics of such anisotropic metamaterials (Fig. 1(c2)) [48-52]. In the limit of the infinite number of ultra-thin layers, such metamaterials can be described within an effective index model by a uniaxial effective dielectric tensor $\hat{\varepsilon} = diag[\varepsilon_{xx}, \varepsilon_{yy}, \varepsilon_{zz}]$. If $\varepsilon_{xx} = \varepsilon_{yy} = \varepsilon_{\parallel}$ and $\varepsilon_{\parallel} \cdot \varepsilon_{zz} < 0$, the metamaterial dispersion relation $\omega^2/c^2 = (k_x^2 + k_y^2)/\varepsilon_{zz} + k_z^2/\varepsilon_{\parallel}$ transforms from that of an ellipsoid to a hyperboloid shown in Fig. 1(c3). This optical topological transition from an ordinary to a so-called hyperbolic metamaterial (HMM) has a dramatic effect on its DOS, as illustrated in Fig. 1(c4). Similar to the case of the SPP on a single interface, the high DOS results from the contribution of the high-momentum states, with the dissipation losses and the finite size of the metamaterial unit cell imposing a cutoff to the largest accessible wavevectors. The presence of multiple high-k branches in the HMM dispersion increases the bandwidth of the high-DOS spectral region over that of the SPP on a single interface (compare Figs. 1(b4) and (c4)). The high-energy limit on this bandwidth is imposed by the metal plasma frequency, and the low-energy one – by the number of metamaterial unit cells and by dissipative losses.

Confinement of SP modes to a surface of a metal nanoparticle (Fig. 1(d1)) further alters their dispersion, only allowing for discrete solutions corresponding to localized surface plasmon modes (Fig. 1(d2)) [8, 9]. These modes have different values of angular momentum characterized by the quantum number $l$, which for a spherical particle of radius $r_{NP}$ has the form $k_\theta = (l(l+1)/r_{NP}^2)^{1/2}$. The spatial field intensity distributions of the dipole ($l = 1$) and quadrupole ($l = 2$) modes are shown in the inset to Fig. 1(d2)). Each LSP mode of a spherical particle has a frequency degeneracy of $2l + 1$, owing to several allowed azimuthal momentum numbers $m = -l,...l$ per each angular number $l$, as illustrated in Fig. 1(d3). This degeneracy is lifted in particles of more complex geometries [9]. Existence of several LSP modes leads to multiple peaks in the LDOS frequency spectrum in the vicinity of nanoparticle (Fig. 1(d4)). The bandwidth of the spectral region where high-LDOS peaks appear is limited by the plasma frequency on the high-energy side, and by the particle size on the low-energy side. In turn, the bandwidth of each peak reflects the mode radiative and dissipative losses.

## 3. Surface-plasmon-enhanced absorbers, emitters and reflectors

The large DOS associated with SP modes opens new horizons in light trapping and emission management. The maximum absorption enhancement in an isotropic medium with index $n$ is restricted by the so-called Yablonovitch's limit of $4n^2$ in the geometrical optics limit [47]. This is a direct consequence of the dispersion relation of photons in bulk (Figs. 1(a2,a3)), which leads to the group velocity in the form $v_g = d\omega/dk = c/n$. As a result, the light intensity (power per unit area) in the material ($I_n$) differs from the light intensity in vacuum ($I_{vac}$) by the factor $n^2$ ( $I_n \equiv U \cdot v_g = n^2 I_{vac}$ ), and the absorption can be enhanced by $4n^2$ (in the presence of a back-reflector). However, reaching this limit requires filling all the available momentum states within the material, and not all of them are directly accessible for coupling to/from the far-field radiation.

This is illustrated in Fig. 2(a), which shows that the light incident on a planar interface of a bulk material can only couple to the continuum of modes within the angular cone $\Omega$. The modes outside of this cone are trapped within the material by the total internal reflection and cannot be directly accessed due to the momentum mismatch with the free-space radiation. If the thickness of the material is on the order of or below the optical wavelength ($h \leq \lambda$), the trapped modes no longer fill a continuum. Instead, they form a set of $M$ quantized guided modes (shown in Fig. 2(b)) with DOS values exceeding that of the bulk material [46, 53].

The momentum states corresponding to the trapped modes in Figs. 2(a,b) can be filled by scattering of the incoming waves by random or engineered perturbations on the material surface or within its bulk region (Fig. 2(c)), which enables exceeding the bulk absorption limit at select frequencies [45, 46, 53, 54]. Although dielectric perturbations can be used for this purpose, metal nanoparticles offer more compact designs and enhanced scattering efficiencies [19, 21]. Likewise, direct coupling from propagating waves into high-momentum SPP modes is not possible, but can be induced by adding localized scatterers such as nanoparticles or gratings (Fig. 2(d)). The large DOS provided by the SPP mode in the vicinity of the interface combined with the DOS of trapped modes increases the light absorption in semiconductors [55]. As the SPP waves do not contribute to the energy density in the far-field, the SPP-mediated absorption enhancement is most pronounced in the limit of very thin absorbers. Excitation of the LSP modes on metal nanoparticles or in nanovoids can also be used to enhance light absorption in the host material owing to the high field intensity generated on particle surfaces (Fig. 2(e)) [17, 19, 21, 56]. Both light polarizations can couple to low-angular momentum LSPs such as dipole modes, and optical coupling between LSPs of two or more particles can be used to excite 'dark' high-momentum LSP modes trapped near the surface of individual particles [10, 57-59].

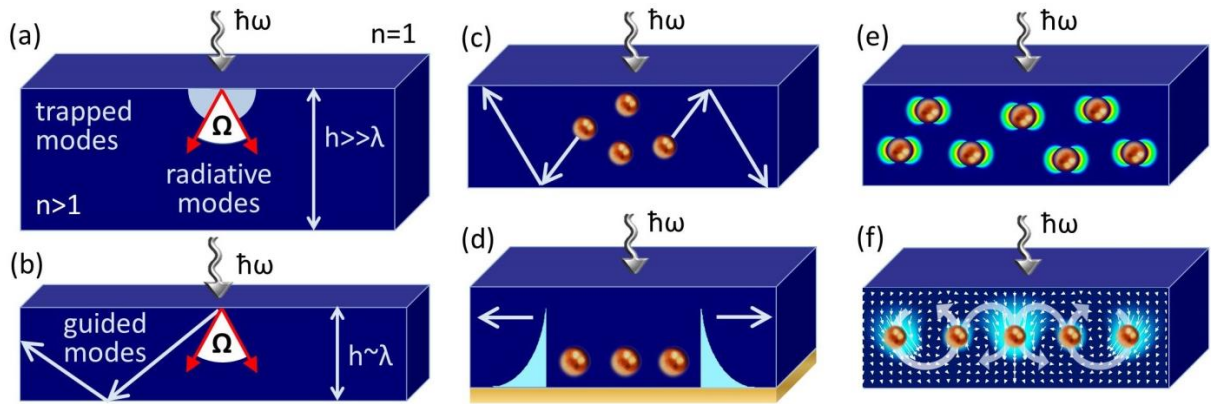

**Figure 2. Surface-plasmon-mediated light trapping and absorption schemes**. Schematics of light trapping in (a) a bulk material and (b) a slab of thickness $h$. The angular cone $\Omega$ (white fill) contains the continuum of waves that couple to the far-field radiation in vacuum, while the modes trapped within the material occupy the range above this cone (blue fill and arrows). (c-f) Mechanisms of SP-mediated absorption enhancement, including: scattering-induced light coupling (c) into the slab guided modes and/or (d) into the SPP mode on the surface on the metal substrate, (e) excitation of LSP resonances on particles or particle clusters, and (f) looping light flow into optical vortices pinned to plasmonic nanostructures.

For PV applications, high dissipative losses in metals may result in most of the energy getting absorbed in metal rather than in semiconductor. This causes significant local heat generation [60], limiting the applications of SP modes for PV energy conversion [61]. How to overcome the dissipative loss of SPs for PV applications remains an active research topic. This problem can be (at least partially) alleviated owing to the ability of plasmonic nanoparticles to strongly modify the phase of the optical field in their vicinity. This makes possible trapping passing photons into optical vortices – nanoscale areas of circulating flow of optical energy around the localized phase singularities in the LSP near field (Fig. 2(f)) [62-64]. If these vortices are engineered to re-circulate optical energy outside of the volume of the particles, metal dissipative losses can be significantly reduced. In addition to the light absorption enhancement in host semiconductors, which results in the generation of photocurrent, plasmonic nanostructures can be used to improve absorption in the upconversion medium and thus to utilize photons with the energies below the semiconductor bandgap [28].

By reciprocity, metal elements can play an important role in the SP-mediated light extraction for applications such as solid-state lighting and coherent thermal emission engineering. The radiative decay rate of an emitter is proportional to the DOS of the free space, but can be enhanced via the Purcell effect [65] if additional high-DOS decay channels are introduced. High-momentum SP modes provide such high-capacity decay channels [13, 23, 48, 49, 66-69]. The radiative rate of a dipole can be enhanced by the Purcell factor, which is proportional to the ratio of the SP-mediated DOS $D_{SP}$ to that of the free-space $D_{FS} = \left(\omega^2/\pi^2 c^3\right)$: $\tau_{SP}^{-1}/\tau_{FS}^{-1} \sim D_{SP}/D_{FS}$. Emission enhancement via the Purcell effect is used for fluorescence sensing or solid-state emission enhancement [13, 23, 24, 68-70], and can also be applied to tailor thermal emission [30, 71]. While the efficiency of coupling into the SP mode can be significantly enhanced, dissipative losses in metals also provide additional channels for non-radiative decay, which may cause emission quenching [68, 69, 72]. To address this issue, plasmonic structures should be engineered to optimize coupling between high-DOS SP modes and radiative waves [23, 69, 70]. Overall, emitters with medium-to-low internal quantum efficiencies benefit most from the SP-mediated radiative rate enhancement, as for them the positive effect of internal non-radiative losses reduction outweighs the negative effect of dissipative losses in metal [68, 72]. An alternative way to strongly enhance the emitter radiative rate and to harvest the emitted light is via evanescent coupling of SP modes to high-DOS trapped modes of dielectric waveguides or optical microcavities [73, 74].

SP-mediated absorbers and emitters for energy applications should provide spectral and angular selectivity within a broad photon energy range (e.g., covering the entire solar spectrum and the thermal emission spectrum). In particular, for TPV and solar-thermal platforms, high absorption in the visible and near-infrared (IR) range should be accompanied by low emission in the mid-to-far IR [75, 76]. The broadband requirement makes SP engineering more challenging than for traditional applications in on-chip communications and bio(chemical) sensing, which require strong optical response at one or few select frequencies. The approaches commonly used to achieve spectrally-selective SP-mediated absorption include: shape and size tuning of individual nanoparticles [77], impedance matching in ordered [78, 79] and random nanostructures [80], Wood anomalies in the scattering from periodic gratings [81-83]. Broadband spectral response can be achieved by using non-resonant mechanisms such as adiabatic focusing and/or Brewster funneling in tapered structures [18, 84-86], by engineering nanoparticles or particle clusters with a broadband response [10, 87, 88], by integrating several plasmonic elements tuned to operate at different wavelengths [78, 89, 90], or by arranging nanoparticles into lattices with quasi-crystalline and aperiodic geometries [11, 23, 91, 92].

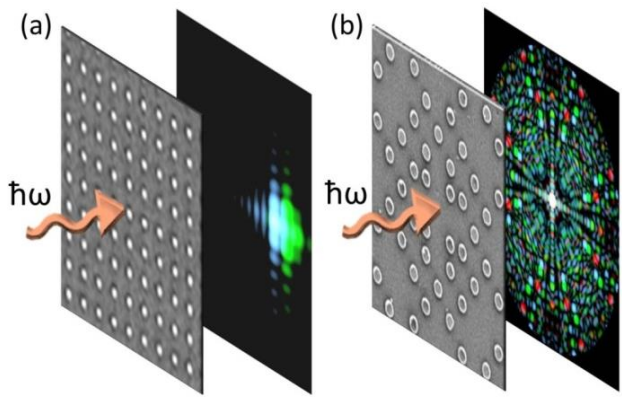

**Figure 3. Angularly-selective scattering by plasmonic arrays.** (a) Periodic arrangements of metal nanostructures offer high angular and spectral selectivity of light absorption and emission by coupling LSP modes to propagating waves with select k-vectors. (b) Multi-scale quasi-periodic or aperiodic arrays can yield broadband and omnidirectional response by providing multiple k-vectors in their reciprocal lattices for coupling to LSP modes [93].

Angular selectivity of the emission is desirable for some applications. By limiting the angular emittance range, it is possible to enhance the collection efficiency of emitted light or to improve the overall conversion efficiency of PV and TPV devices [94, 95]. In particular, the absorption limit of isotropic semiconductor material can be increased to $4n^2/\sin^2\theta$ if its surface emission/absorption is limited to a cone with half-angle $\theta$ [46, 47]. As illustrated in Fig. 3, far-field constructive interference of light scattered by plasmonic elements arranged into periodic structures enhances light absorption or extraction within a certain angular range [70, 81, 82, 93, 96]. On the other hand, complex aperiodic plasmonic lattices may be engineered to provide omnidirectional light absorption or emission [11, 23, 91, 92]. Finally, near-field destructive interference effects can be used to tailor angular emission characteristics of plasmonic nanostructures [62, 67, 97].

Strong near-field interactions between plasmonic nanoparticles can also be exploited for the design of dynamically tunable 'smart' mirrors and coatings. Individual particles serve as local scattering centers for photons having energies overlapping with the particle LSP resonances. The light is preferentially scattered into the high-index material of the absorber rather than back into the air region owing to the $n^2$ difference in the DOS between the absorber and the air. This effect can be used to create plasmonic antireflection coatings for solar harvesting applications [21, 79]. However, even a monolayer of self-assembled densely-packed nanoparticles forms a very efficient mirror, which reflects photons. This effect can be used to dynamically and reversibly switch the reflection properties of materials and surfaces by moving metal particles towards or away from the surface by applied voltage [42]. This is illustrated in Fig. 4(a), which contrasts the reflection characteristics of Ag nanoparticles dispersed in an electrolyte and of a monolayer of the same particles self-assembled on the interface between two transparent electrolytes. In the case of low interface coverage, visible light propagates through with minor absorption as the LSP resonances of small particles lie outside the visible band. However, the reflection band of the monolayer is significantly red-shifted (Fig. 4(a)) owing to electromagnetic coupling between particle LSP modes (Fig. 4(b)) [10], creating a mirror that blocks the sunlight.

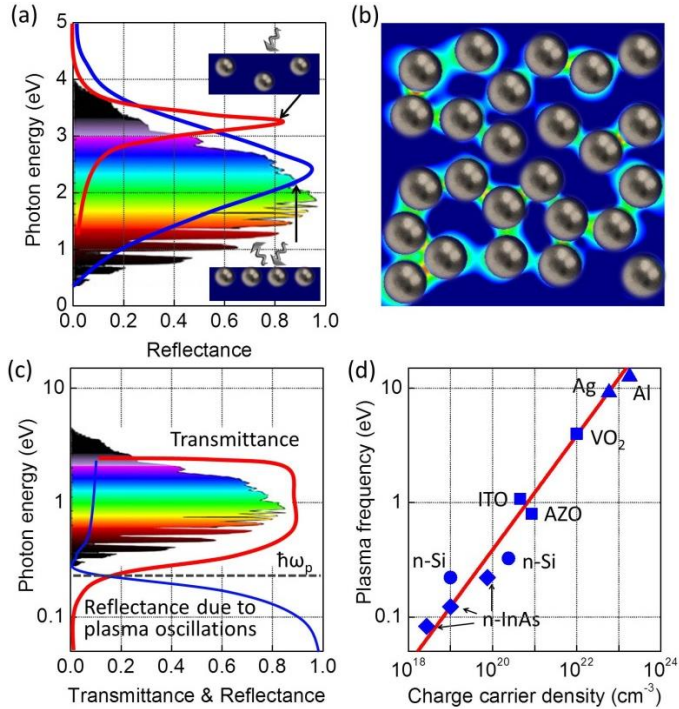

**Figure 4. Smart mirror designs exploiting surface plasmon effects**. (a) Frequency dependence of the reflection coefficient of a liquid-liquid interface covered by nanoparticles. Red line: 33% coverage (top inset), blue line: 100% interface coverage (bottom inset) [42]. (b) An electric field intensity distribution around nanoparticles randomly aggregated on a surface. The near-field coupling between particles re-shapes the reflection spectrum as illustrated in (a). (c) Typical optical transmission and reflection spectra of a film made of a material with plasma frequency $\omega_p$. (d) Plasma frequency as a function of the charge carrier density $n_c$. Red line: functional relation $\omega_p \sim \sqrt{n_c}$, blue dots: plasma frequency of select materials, triangles – noble metals, squares – metal oxides, circles – doped Silicon, diamonds – doped Indium Arsenide [41, 98-103].

On the other hand, the hybrid nature of SPP waves can be used to engineer reversible mirrors that efficiently block the infrared portion of the solar spectrum without affecting the visible light transmission. Although volume oscillations of electrons dramatically restrict transparency of plasmonic materials at frequencies lower than their plasma frequencies, some materials such as e.g. metal oxides and chalcogenides remain highly transparent for higher-frequency photons (Fig. 4(c)) [41, 98, 99]. The cut-off plasma frequency is defined by the concentration of free electrons in the material: $\omega_p^2 = n_c e^2 / m_e \varepsilon_0$ ($n_c$ is the charge carrier density, $e$ is the charge of the electron, $\varepsilon_0$ is the vacuum permittivity, and $m_e$ is the effective electron mass). The electron density can be controlled irreversibly by doping the material and reversibly by the applied voltage. For example, the transmission and reflection spectra of a smart window made of Tin-doped $In_2O_3$ (ITO) nanocrystals can be dynamically switched via electrochemical modulation of the ITO electron concentration, which enables to control the amount of transmitted near-infrared radiation [41]. Fig. 4(d) shows the plasma frequencies of different types of plasmonic materials, including metals, metal oxides, and doped semiconductors as a function of their electron concentration [41, 98-103]. Clearly, materials with plasma frequencies within a wide frequency range from ultra-violet to mid-infrared can be found or synthesized, paving the way to their applications as smart reflectors.

## 4. Plasmon-mediated near-field heat transfer and non-equilibrium processes

As discussed in the previous section, although high LDOS exists near metal surfaces, the far field emission extraction is limited by the efficiency of coupling to the low-DOS free space modes. For example, even though the thermal emittance of individual nanoparticles can be larger than one, thermal emission averaged over an area much larger than the wavelength is less than one even if it contains multiple SP structures. The high LDOS near SPP surfaces decays exponentially over the distance of the order of wavelength. However, when two surfaces are close, near-field energy transfer between surfaces can be achieved via the tunneling process – tunneling of photons rather than electrons – as long as the separation between surfaces is much larger than the electron de-Broglie wavelength [30, 31, 104-106]. Thermally-induced charge fluctuations caused by the chaotic thermal motion of electrons in the material can excite SPP modes, making possible heat exchange between two surfaces via evanescently-coupled SPP waves as schematically illustrated in Fig. 5(a). Total heat flux between two surfaces depends on the number of channels available for the energy transfer in both photon energy and momentum space – i.e., on the LDOS inside the gap – while the capacity of each channel is governed by the hot-side temperature [107]. As a result, the radiative heat exchange between surfaces separated by nanoscale gaps is no longer governed by the Planck's law of thermal radiation, and can be resonantly increased by orders of magnitude as shown in Fig. 5(b). Analogous effect of the tunneling of surface phonon-polaritons have been demonstrated experimentally, showing 3 orders of magnitude higher heat flux than the blackbody radiation when two surfaces are separated below 100 nm [20, 31].

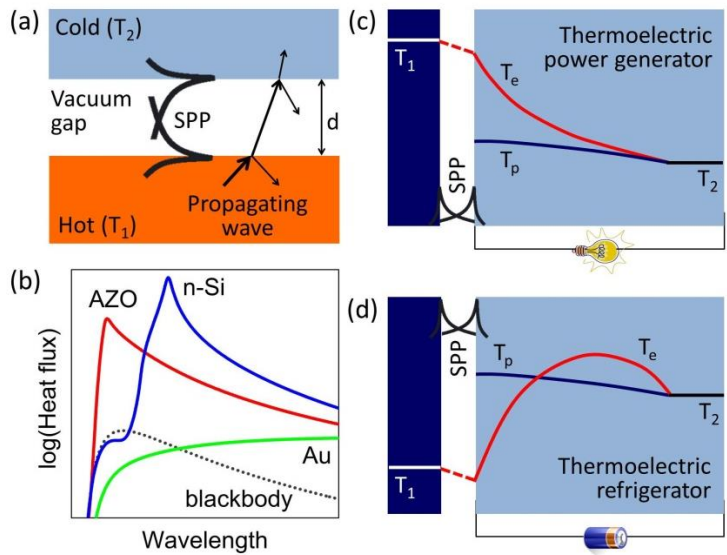

**Figure 5. SPP-mediated heat transfer and thermoelectric energy generation**. (a) Heat transfer between hot and cold plates is mediated by both propagating waves and evanescently-coupled SPP modes. (b) The SPP-enhanced heat flux between surfaces of various plasmonic materials including gold (Au), doped silicon (n-Si), and aluminum-doped zinc oxide (AZO). (c-d) Schematics of the coupled-SPP non-equilibrium thermoelectric devices: (c) a power generator, and (d) a refrigerator [29].

For appreciable photon tunneling to happen between two surfaces at temperatures below 2000K, the SPP wavelength should fall in the infrared, overlapping the peak region of the blackbody spectrum with the central wavelength of $\lambda_T = b \cdot T^{-1}$ ($b = 2.9 \cdot 10^{-3} m \cdot K$), according to the Wien's displacement law, although the peak thermal emission from SPP waves may not overlap exactly with this law. Traditional plasmonic materials such as Au have plasma frequency too far in the ultraviolet to provide good overlap with the thermal emission spectrum. However, new plasmonic materials, including metal oxides and nitrides [98, 99], graphene [36, 108], and highly doped semiconductors such as Si [101, 109, 110], InSb and InAs [102], can provide strong resonant enhancement of the near-field heat flux (Fig. 5(b)). Hyperbolic metamaterials can also be used for a broadband enhanced near-field radiation transfer [111].

Another idea of using the near-field coupling of SPP waves for energy conversion exploits the fact that, unlike other types of surfaces waves, SPPs as are hybrid waves resulting from coupling of electromagnetic field with coherent oscillations of free electrons. Accordingly, coupled SPPs can be used to create hot or cold electrons for thermoelectric power conversion or refrigeration, as shown in Fig. 5(c,d) [29]. For example, the high heat-induced energy of the SPP mode on the hot surface at temperature $T_1$ can be imparted directly to the electrons on the cold surface (at $T_2$) across a nanoscale vacuum gap, without physical electron tunneling (Fig. 5(c)). This creates non-equilibrium between the electrons and phonons on the cold side, whose energies are now governed by the Fermi-Dirac and Bose-Einstein distribution functions at different temperatures:
$f_{FD}(E,T_e) = \left(\exp\left((E-E_F)/k_B T_e\right) + 1\right)^{-1}$ for electrons ($E_F$ is the Fermi level), and
$f_{BE}(E,T_p) = \left(\exp\left(E/k_B T_p\right) - 1\right)^{-1}$ for phonons. The non-equilibrium temperature distribution can extend over the distances defined by the electron-phonon cooling length, which in different materials varies from 100nm to several microns. As in thermoelectric devices only the electrons do the useful work, and the phonon-mediated heat conduction hinders the energy conversion efficiency, the non-equilibrium devices shown in Fig. 5(c,d) can potentially outperform classical ones [29].

## 5. Hot electrons generation and plasmon-enhanced photocatalysis

The hybrid nature of surface plasmons can also provide additional pathways for harvesting the energy of below-gap photons in PV platforms, via hot electron tunneling from metal to semiconductor as shown in Fig. 6(a) [32-35, 112]. Energy of photons incident on the metal surface drives the resonant collective electron oscillations and rises the energy of electrons above the Fermi level to higher energy levels, creating so-called hot electrons (Fig. 6(b)). Normally, these hot electrons cool very fast due to scattering by phonons, on picosecond timescale in most metals [44]. If the hot electrons can be extracted out before they cool down, higher energy conversion efficiency can be achieved. One scheme is to transfer hot electrons in the metal volume to a semiconductor by tunneling through a Schottky barrier (Fig. 6(c)). SP resonances of metal nanostructures can be tuned across the visible and IR spectral ranges by the proper choice of the material and geometry. As such, hot electrons can be generated by free-carrier absorption of the photons with the energy below the semiconductor bandgap (Fig. 6(c)), which offers the way to increase the conversion efficiency of PV cells and to extend the bandwidth of photon detectors.

The concentration of hot electrons in metals can be controlled by the LDOS of the SP resonance. Efficient coupling of incoming photons into higher-LDOS modes such as SPP waves can increase the hot electrons concentration. This illustrated in Fig. 6(d), which shows a metal-insulator-metal device that generates a photocurrent due to light absorption and tunneling of hot electrons across the insulator [113]. Under direct light illumination of both metal electrodes, currents of opposite signs cancel each other. However, by increasing the efficiency of light-to-SPP coupling on one of the electrodes (e.g., by using a prism coupler), a photocurrent can be generated by tunneling to the other side across the barrier, as long as this tunneling is faster than the phonon cooling.

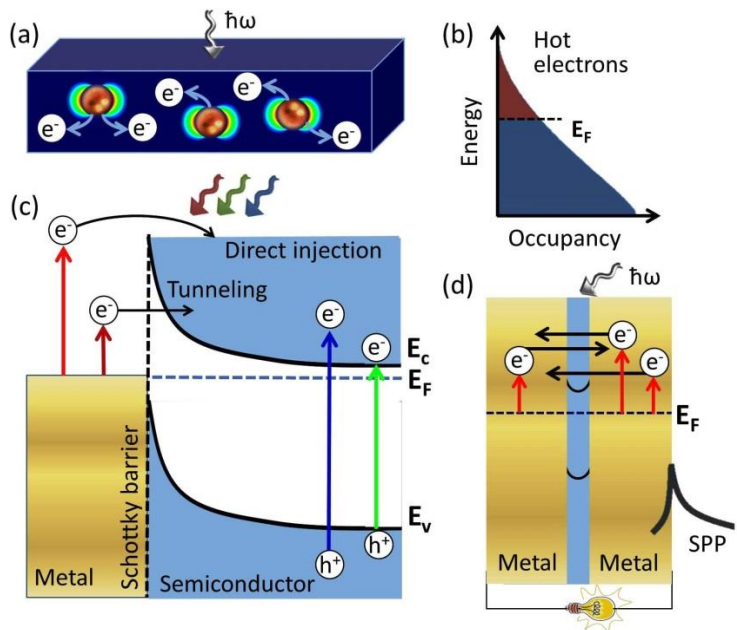

**Figure 6. SP-mediated hot-electron generation and harvesting**. (a) A schematic of the hot electron injection from metal nanoparticles into a host semiconductor. (b) SP-induced energy distribution of the electrons in metal. (c) Photocurrent generation in metal-semiconductor PV devices via absorption of above- and below-gap photons. Below-gap photons generate hot electrons in metal, which can tunnel through the Schottky barrier into the semiconductor and contribute to the photocurrent [32, 34, 35]. (d) Hot-electron-driven photocurrent generation in the metal-insulator-metal device [113].

SP modes excited by the photon absorption can also be exploited to drive chemical reactions, which makes plasmonic nanostructures promising platforms for photocatalysis [37, 114-118]. It has been demonstrated that plasmonic nanostructures can drive catalytic reactions at lower temperatures than their conventional counterparts that use only thermal stimulus because they enable channeling of light energy into chemical reactions in a directed, orbital-specific manner. For example, SP-generated hot electrons can populate anti-bonding orbitals of $O_2$, $H_2$ and $D_2$ molecules adsorbed on the metal surface, which facilitates the dissociation reactions [37, 118]. A combination of thermal and hot-electron effects provided by SP modes can yield orders-of-magnitude enhancement of catalytic reaction rates [114].

The ability to tune the frequency of SP modes across a broad range can be used to drive catalytic reaction under visible or IR light illumination. For example, photocatalytic production of hydrogen from water is typically achieved via the Honda-Fujishima effect [119], which makes use of the electrons and holes generated in $TiO_2$ (titanium oxide) semiconductor electrode under illumination by the ultraviolet light. As illustrated in Fig. 7(a), integration of Au nanoparticles on the surface of $TiO_2$ electrodes also enables the use of the visible part of the solar radiation and thus improves the reaction efficiency. Furthermore, free electrons generated in $TiO_2$ are efficiently captured by Au particles, which prevents their recombination in the semiconductor volume and further enhances the reaction efficiency [40]. Recently, it has been demonstrated that it is possible to engineer photocatalytic devices with all the carriers generated by the SP-mediated electron heating in metals rather than by electron-hole generation in semiconductors. One possible realization of such a device is shown in Fig. 7(b) [38, 39], where hot electron generated in Au by the visible light are transferred to a platinum catalyst, which drives the reduction reaction. In turn, a cobalt catalyst helps to drive the oxidation reaction by helping Au particle to capture electrons from water.

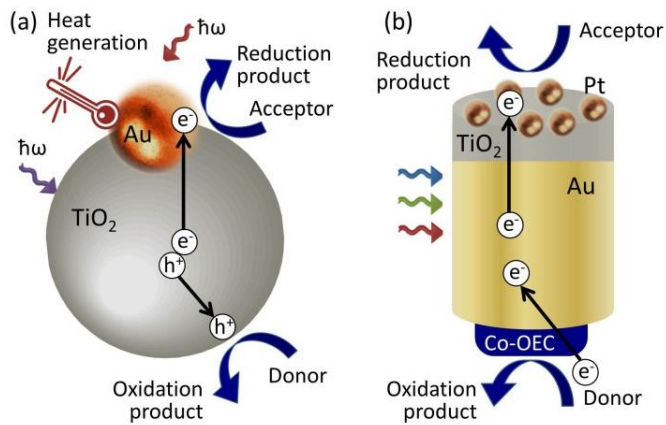

**Figure 7. Surface-plasmon-driven photo-catalysis**. (a,b) Schematics of the SP-mediated photo-catalytic devices. (a) A hybrid device exploiting electrons generated both in semiconductor electrodes and on metal nanoparticles [40]. (b) Autonomous solar water-splitting device in which most electrons originate from the excitation of LSP modes on Au nanoparticles [38, 39].

Finally, SP resonances can be used to not only drive but also to non-invasively monitor catalytic processes. As the high-LDOS SP resonances are very sensitive to the presence of molecules adsorbed on the metal surface, the frequency shift of the SP resonance can be used to detect the adsorption of molecular species that bring the catalytic reaction to a halt [120].

## 6. Surface-plasmon-induced local heating and phase transition

In addition to applications in photocatalysis, SP-induced heating of nanoparticles has already been used to induce phase transformations in various materials [60, 121, 122]. One application of the local heating effect is to kill cancer tumors. Melting of plasmonic nanoparticles themselves have been observed when the incident laser wavelength is tuned to the LSP frequency [123], which can be useful for optical data storage applications [43]. Another example is recently reported undercooled boiling of water under concentrated sunlight when it is seeded with nanoparticles (Fig.

8(a)) [25, 27]. Although vapor bubbles were generated, the water was still close to ambient temperature. Similarly, it was reported that SP-induced heating generates vapor over the localized regions of the catalytic sites, while keeping fluids and substrates at room temperature [115]. The puzzle is that the theoretical estimate of the nanoparticles temperature rise in [25], which is based on the calculation of the absorbed power and the solution of the Fourier law heat conduction equation, is only 0.04°C.

For sub-wavelength-size nanoparticles, the classical model predicts that the surface temperature rise is proportional to the square of the particle radius, $\Delta T \sim r_{NP}^2$ [121]. Fig. 8(b) shows the measured threshold temperatures for vapor formation on Au particles of various sizes reported in the literature (symbols) and compares them with theoretical predictions (lines). Only in one case the measured threshold is close to the model prediction. The comparison suggests that the fundamental understanding of energy flux for phase transformation at this scale is still unknown. The nature of the heat transfer between the nanoparticle and vapor/liquid phases of water at the nanoscale (i.e., diffusive or ballistic), the mechanism of vapor phase nucleation, the influence of pressure field on heat transfer, the growth kinetics of vapor phase, vapor-induced particle buoyancy, fluctuations of absorbed energy due to formation of vapor phase and radiation between the phases are among the topics that needs to be explored to gain a better understanding of this phenomenon (Fig. 8(a)).

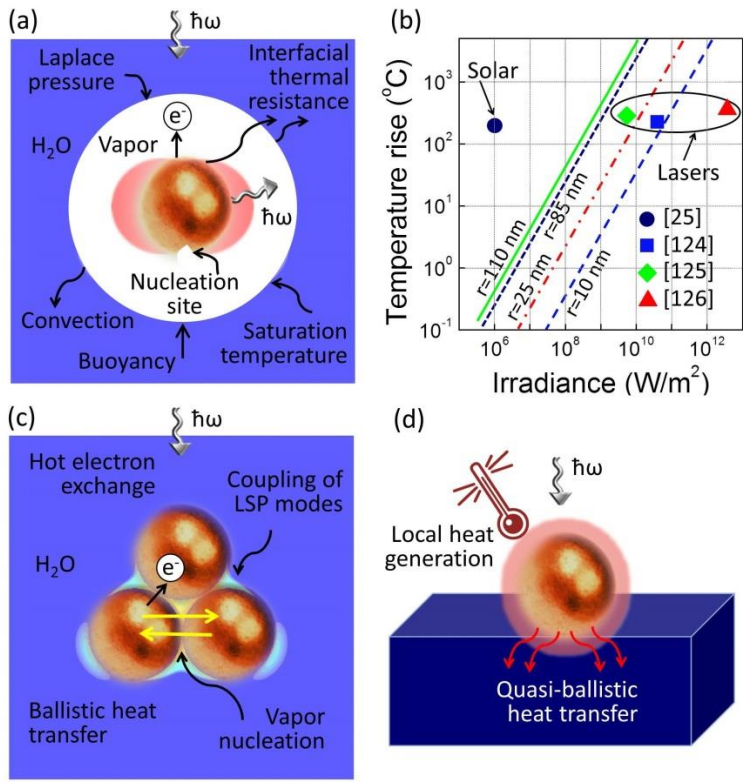

**Figure 8. SP-induced phase transitions.** (a) Complex dynamics of the localized phase transition around the metal nanoparticle heated via the LSP excitation by incoming photons. (b) Experimental (symbols) [25, 124-126] and theoretical (lines) values of the surface temperature rise on Au nanoparticles of different radii (shown as labels). Lines and dots of matching colors correspond to the same nanoparticle size. (c) Basic processes underlying optical and thermal interactions within multi-particle clusters. (d) Localized heating of nanoparticles smaller than the phonon mean-free path in the substrate material.

For ensembles of coupled nanoparticles (Fig. 8(c)), additional effects may play a role and further complicate the picture [26, 121, 127, 128]. These include near-field SP coupling that creates localized electric field hot-spots within the cluster as well as the heat and hot electron exchange between adjacent nanoparticles. Theoretical modeling predicts the monotonic temperature increase with the decrease of the inter-particle gap [128], and non-uniform temperature distributions around the nanoparticles that follows the distribution of the hot-spots [25].

Localized heating and phase transformations can also be induced by the SP modes excited on particles immobilized on substrates (Fig. 8(d)). If the particle size is smaller than the phonon mean-free path in the substrate material, the heat transfer from the particle is significantly reduced [129]. This makes possible local heating of particles [130], creating localized areas of vapor [26] or super-heated water around them [125], and even localized melting or evaporation of the substrate material [131]. Finally, SP-assisted local phase transformations can be used to convert the energy of absorbed photons into mechanical energy to drive the flow of liquids in microfluidic channels or to fabricate multi-particle plasmonic nanostructures [124, 132].

## 8. Discussion and outlook

As demonstrated in the previous section, SP effects on metal surfaces and nanoparticles offer many advantages for renewable energy applications. These include extreme energy concentration within sub-wavelength volumes, which enables miniaturization of light harvesting platforms, and the ability to generate hot electrons, which opens new pathways for solar-to-electrical and solar-to-chemical energy conversion.

A limiting factor in using metal structures for light trapping is the high spatial localization of the SP-modified DOS, ranging from tens to hundreds of nanometers from metal surfaces. Accordingly, the performance of thick PV cells can primarily be boosted via the SP-induced scattering, which has been successfully utilized to engineer plasmonic anti-reflection coatings [133, 134]. Plasmonic coatings combined with the standard interference coatings yielded up to 9% improvement in light transmission over the interference coatings alone [79, 134, 135]. Another important performance improvement in the PV technology fueled by plasmonic scattering effects has been in the significant shrinkage of PV cells without degrading their efficiency. For example, reduction of the Si absorbing layer thickness from over a hundred to a just few micron helps to significantly reduce the material cost per output power [21].

The thin-film cells made of materials with low minority carrier diffusion lengths such as polycrystalline [134-136] and organic [137-139] semiconductors or quantum dots [140-142] as well as cells with up- or down-converters [28] are expected to benefit most from both the enhanced scattering and the SP-modified localized high DOS. Indeed, over 20% efficiency improvement has been achieved for thin-film a-Si PV cells owing to enhanced light scattering from metal nanoparticles on the surface of the active layer [135, 136]. However, the biggest efficiency boosts of thin-film polymer PV cells reported to date stem from the localized near-fields generated by plasmonic nanoparticles embedded into the active layer. Although the highest achieved efficiency of the SP-enhanced polymer cell is only 8.92% [137], the reported efficiency improvements of various plasmonic over non-plasmonic polymer cells are as high as 200% [138, 139]. Such improvements pave the way to significantly surpassing the 10% overall cell efficiency limit, which is necessary to make polymer cells truly competitive in the marketplace [138, 139].

Another downside of using metal structures for light trapping and generation is their high level of dissipative losses, which hinders the performance of SP-enhanced PV cells [61] and SP-enhanced emitters [23]. Accordingly, the scattering and field enhancement effects should be strong enough to overcome the parasitic absorption in the metal. Metal inclusions in PV active layers can also act as recombination centers, leading to additional efficiency losses. One possible approach to alleviate both sources of losses is to add thin insulating layers between plasmonic structures and active PV layers [143] or thin insulating shells enclosing metal nanoparticle cores [137, 141]. Further exploration of plasmonic structures that maximize energy recycling outside of the metal volume [62-64, 144] is also expected to significantly reduce the dissipation losses and to improve performance of PV cells and SP-enhanced solid-state light sources.

On the other hand, other energy-related applications of plasmonic materials such as SP-mediated heat generation and enhanced near-field heat transfer can benefit from using plasmonic materials with both low and high dissipative losses. In particular, the efficiency of a near-field TPV system is maximized when a low-loss metal surface supporting a high-DOS narrowband SPP mode is used as the quasi-monochromatic source of photons with the energies slightly exceeding the PV cell bandgap [22]. Replacing a conventional far-field blackbody source with a quasi-monochromatic near-field source at the same temperature can significantly increase the TPV system overall efficiency. For example, the use of the optimized SPP source at T=2000K and 5nm vacuum gap is predicted to increase the efficiency of the TPV system with a GaSb cell from 13% to 35% [22]. This efficiency exceeds the thermodynamic limit of 29% for the GaSb cell illuminated by the far-field source at the same temperature. However, even a non-optimized broadband near-field SPP mode supported by a tungsten-vacuum surface is estimated to provide high TPV conversion efficiency of 27% at the same conditions [22]. High tunability of the SP resonance spectral position by the electron concentration in the material and by the structure geometry enables creation of coherent near-field and far-field thermal sources [22, 71, 101, 110, 145] compatible with various types of PV cells. Reaching nm-scale vacuum gap widths represents a major technological challenge; however, the SP-mediated efficiency increase has been demonstrated experimentally for the TPV systems with the gaps in the 500 nm-1 micron range [146, 147].

For the applications other than the near-field TPV energy conversion, such as e.g. near-field cooling or heat-assisted recording, the total useful near-field heat flux between hot and cold surfaces is calculated by integrating over all the available channels in the energy and momentum space. Although the dissipation-induced damping of SPP modes reduces their maximum accessible k-vectors (Figs. 1(b2,c2)), it simultaneously increases their frequency bandwidth [110], which may lead to the overall increase in the energy transfer rate. This makes the near-field heat transfer an ideal niche application for plasmonic materials with high dissipative losses. The use of metamaterial surfaces can further increase the frequency bandwidth and thus the total heat flux [111].

Strong dissipation-induced localized heating of nanoparticles can also be extremely useful for vapor generation, water distillation, and photocatalysis [25, 27, 40, 115]. For example, hot working fluids or superheated steam are currently generated by using parabolic troughs and power towers. The steam can later be used to generate electricity with efficiency of 35-40%, although the levellized cost of energy for solar-thermal processes is higher than for PV approaches [148]. Conventional vapor-generating systems are not portable and require high levels of optical concentration and high temperature working fluids and storage systems This makes photo-thermal conversion with nanoparticles [25, 27, 60, 122] an attractive alternative route to engineering low-cost portable superheated steam generators. Although these plasmonic approaches are still at the research level, the recently demonstrated solar-to-hot vapor conversion efficiency of up to 24% [25, 27] is a good indication that they may emerge as a competitive solar-thermal conversion technique in the not-so-distant future. An open issue in this research is a still-missing theoretical understanding of the complex photo-thermal processes on nanoparticles and the quasi-ballistic heat transfer away from or between the particles and/or surfaces (Fig. 8) [25, 27, 128, 129].

Finally, further technological improvements are needed to boost low efficiencies of the devices based on harvesting the SP-induced hot-electrons [38, 39]. Recent theoretical research predicts that the maximum efficiency limit of the energy converters based on the enhanced internal photoemission in metal-semiconductor Schottky junctions is below 8% even if perfect optical absorption can be achieved using plasmonic nanostructures. This limit stems from the fundamental electronic properties of metallic absorbers. However, modification of the electron density of states of the plasmonic material of the absorber offers the way to increase the efficiency above 20% level [133]. Overall, the major role of plasmonic effects in the performance enhancement of hot-electron PV cells and artificial photosynthetic systems lies in their ability to harvest near-infrared and infrared portions of the solar spectrum, which cannot be utilized in the conventional PV and photo-catalytic devices.

## Acknowledgments

This work has been supported by the 'Solid State Solar-Thermal Energy Conversion Center (S3TEC)', funded by the US Department of Energy, Office of Science, and Office of Basic Energy, Award No. DE-SC0001299/DE-FG02-09ER46577 (for solar thermal applications), Grant no. DE-FG02-02ER45977 (for near-field transport), DOE SunShot Grant no. 6924527 (for photovoltaic applications), and by AFOSR MURI Grant no. UIUC FA9550-08-1-0407 (for thermal transport in metamaterials).